\documentclass{moriond}

\def\Journal#1#2#3#4{{#1} {\bf #2}, #3 (#4)}

\def\PLB{{\em Phys. Lett.}}
\def\PRL{\em Phys. Rev. Lett.}
\def\PRD{{\em Phys. Rev.}}

\def\JHEP{{\em JHEP}}
\def\EPJ{{\em Eur. Phys. J.}}

\def\be{\begin{equation}}
\def\ee{\end{equation}}
\def\bea{\begin{eqnarray}}
\def\eea{\end{eqnarray}}

\DeclareRobustCommand{\nnlojet}{NNLO\scalebox{0.8}{JET}\ }

\newcommand{\Photo}{}

\begin{document}
\vspace*{4cm}
\title{ISOLATED PHOTON AND PHOTON+JET PRODUCTION AT NNLO QCD ACCURACY AND THE RATIO $R_{13/8}^\gamma$}

\author{X. CHEN,$^a$ T. GEHRMANN,$^a$ N. GLOVER,$^b$ M. H\"OFER,$^a$ A. HUSS$^c$}

\address{$^a$ Physik-Institut, Universit\"at Z\"urich, Winterthurerstrasse 190, CH-8057 Z\"urich, Switzerland\\
	$^b$ Institute for Particle Physics Phenomenology, Durham University, Durham, DH1 3LE, UK\\
	$^c$ Theoretical Physics Department, CERN, CH-1211 Geneva 23, Switzerland}

\maketitle\abstracts{
	We discuss different approaches to photon isolation in fixed-order calculations and present a new next-to-next-to-leading order (NNLO) QCD calculation of $R_{13/8}^\gamma$, the ratio of the inclusive isolated photon cross section at $8~\mathrm{TeV}$ and $13~\mathrm{TeV}$, differential in the photon transverse momentum, which was recently measured by the ATLAS collaboration.
	}

\section{Introduction}
	Both inclusive isolated photon ($\gamma+X$) and photon-plus-jet ($\gamma+j$) production in $pp$ collisions present a means to test QCD dynamics using a colourless probe. Because their Born-level processes are $\bar{q}q\rightarrow g\gamma$ and $qg\rightarrow q\gamma$, related observables are sensitive to the gluon-distribution in the proton already at leading order (LO).
	
	Recent experimental analyses by ATLAS~\cite{ATLAS,ATLAS2} and CMS~\cite{CMS} pushed the experimental uncertainties down to a few percent. To match this accuracy also in theory calculations, the inclusion of NNLO QCD corrections is crucial. They have been calculated for $\gamma+X$ and $\gamma+j$ at $\sqrt{s}=8~\mathrm{TeV}$ by the MCFM~\cite{MCFM} collaboration. In our recent paper~\cite{paper} we present an independent calculation of the NNLO corrections, using the \nnlojet framework. \nnlojet is a parton-level event generator which uses the antenna subtraction method~\cite{Antenna} to subtract the infrared (IR) QCD divergencies. The matrix elements for $\gamma+X$ and $\gamma+j$ are implemented up to NNLO in analytic form.	

	In the experimental environment it is necessary to separate any photon produced in the hard partonic scattering process from photons of other origin, for example radiation occurring during the hadronization process. One therefore measures the hadronic energy in the vicinity of the photon and defines conditions for its shape and amount. If these are met, the photon is said to be isolated. 
	
	When reconstructing the experimental isolation procedure in fixed-order theory calculations, one has to deal with hadronic radiation arbitrarily collinear to the photon. This must be taken care of by either including photon fragmentation functions to the perturbative order under consideration, which to NNLO has not been done so far, or by modifying the isolation prescription to eliminate the collinear configurations. In the latter approach a systematic difference between isolation procedures used in experiment and theory emerges.   

\section{Photon Isolation}
	There are several prescriptions for the photon isolation. They mainly differ in how exactly the "vicinity" of the photon is defined and how the hadronic energy therein may be distributed. The two most common ones are the fixed (hard) cone isolation, and the dynamical cone (Frixione~\cite{Frixione}) isolation.
		
	\paragraph{Fixed cone isolation - } A cone around the photon axis is defined by the distance $R=\sqrt{\Delta\eta^2+\Delta\phi^2}$, called the radius of the cone. The integrated hadronic transverse energy within the cone has to be smaller as a certain $E_T^\mathrm{max}$ for the photon to be considered as isolated. Often $E_T^\mathrm{max}$ is given as a simple linear function of the photon transverse momentum/ energy:
		\begin{equation}
			E_T^\mathrm{max} = \varepsilon E_T^\gamma + E_T^\mathrm{thres}\,. 
		\end{equation}
	This isolation criterion is used in all experimental analyses so far. It allows, however, for hadronic radiation arbitrarily collinear to the photon, as long as its energy is not to large. This introduces a sensitivity to the photon fragmentation, which is difficult to describe from the theoretical viewpoint. On the other hand it is not possible to simply set $E_T^\mathrm{max}=0$, because this would, while indeed eliminating the fragmentation sensitivity, cut out part of the soft phase space, rendering observables IR unsafe.		
				
	\paragraph{Dynamical cone isolation - } Instead of a fixed $E_T^\mathrm{max}$ one defines a profile $E_T^\mathrm{max}(r_d)$ with $E_T^\mathrm{max}(r_d)\rightarrow0$ as $r_d\rightarrow 0$. $r_d$ is again the distance from the photon. For any sub-cone with $r_d$ smaller than some maximal radius $R_d$ the integrated energy within this sub-cone must not exceed $E_T^\mathrm{max}(r_d)$. The functional form of the profile conventionally used is
		\begin{equation}
			E_T^\mathrm{max} = \varepsilon_dE_T^\gamma\left(\frac{1-\cos r_d}{1-\cos R_d}\right)^n \qquad \mathrm{for\ all\ \ } r_d<R_d\,. 
		\end{equation}
		This prescription both eliminates the fragmentation sensitivity and ensures IR safety. It can, however, only be approximated in experiments and so one has to tune the parameters of the dynamical isolation to fit the experimental setup as closely as possible.\\		
		
	This difference in the isolation procedures used in experiment and theory is unsatisfactory, as it is a source of uncertainty, which is difficult to quantify. Only the inclusion of the photon fragmentation functions to the same order as the partonic calculation can solve this issue. To NNLO this has not been done so far. But an improvement over the current situation can already be achieved by combining both fixed and dynamical cone in a hybrid approach~\cite{Hybrid}, as used by ATLAS in their $\gamma+j$ study~\cite{ATLAS2}.
	
	\paragraph{Hybrid cone isolation - } A dynamical cone with comparatively small $R_d$ is used to eliminate the fragmentation contribution. In a second stage of the isolation a fixed cone with $R^2\gg R_d^2$ is applied, the parameters of which are chosen according to any experimental analysis under consideration. In this way observables should retain the correct dependence on the parameters of the outer "physical" isolation cone. A residual dependence on the inner dynamical cone remains, but can in principle be made small for a suitable choice of parameters. In our paper~\cite{paper} we present some technical studies on the choice of the inner cone parameters. We calculated the total cross section for $\gamma+X$ at $13~\mathrm{TeV}$ as a function of the outer isolation cone radius $R$, too. It would be interesting to see this analysis performed also in experiment.  	

\begin{figure}[p]\label{fig:Ratio}
	\includegraphics[width=\textwidth]{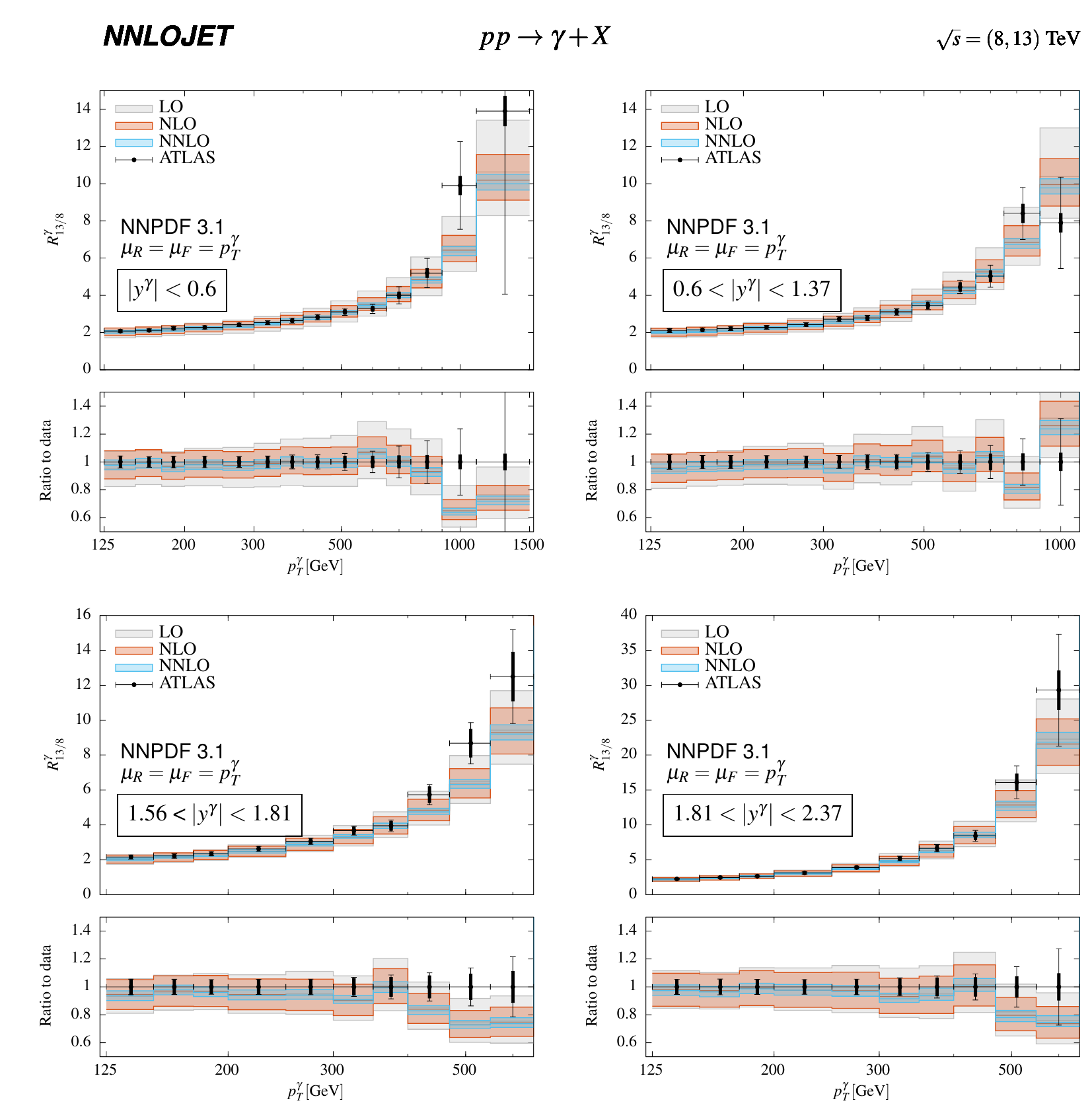}
	\caption{$R^\gamma_{13/8}$ as a function of the transverse energy/momentum of the isolated photon at LO, NLO and NNLO in four different rapidity bins, from central (top left) to most forward (bottom right). The theoretical uncertainty bands are derived by means of an independent variation of factorization and renormalization scales, both in the numerator and the denominator (see text for details). The results are compared to ATLAS data~\protect\cite{ATLAS2019}.}
\end{figure}

\section{The ratio $R_{13/8}^\gamma$}
	In our paper~\cite{paper} we calculated, using the hybrid isolation procedure, several differential distributions for $\gamma+X$ and $\gamma+j$ at $8~\mathrm{TeV}$ and $13~\mathrm{TeV}$, based on studies by ATLAS~\cite{ATLAS,ATLAS2} and CMS~\cite{CMS}. We found that the inclusion of NNLO corrections leads to an significant improvement in both the accuracy of the predictions and the description of the data. Amounting to no more than a few percent, the theory uncertainties are now competitive with experimental errors.  

	Here we present a NNLO calculation of the ratio $R_{13/8}^\gamma$ of the $\gamma+X$ cross section at $8~\mathrm{TeV}$ and $13~\mathrm{TeV}$, differential in $p_T^\gamma$ and presented in four rapidity bins. It is based on a recent measurement of this quantity by ATLAS~\cite{ATLAS2019}. Measuring ratios is a means to reduce the experimental systematic uncertainties.
	
	Both the $8~\mathrm{TeV}$ and the $13~\mathrm{TeV}$ measurements of the $p_T^\gamma$-distribution in isolated photon production by ATLAS were performed in four different regions in rapidity
		\begin{equation}
			|y^\gamma|<0.6\,,\qquad 0.6<|y^\gamma|<1.37\,,\qquad 1.56<|y^\gamma|<1.81\,,\qquad 1.81<|y^\gamma|<2.37\,,
		\end{equation}
	which excludes the region $[1.37,1.56]$. The ratio is measured in the same bins, using the overlap of the phase-space regions of both measurements, with $p_T^\gamma>125~\mathrm{GeV}$. 
	
	For the \nnlojet prediction we use the NNPDF3.1 PDF set and a hybrid photon isolation with parameters
		\begin{equation}
			\begin{array}{lll}
				R_d = 0.1\,,& \qquad\varepsilon_d = 0.1\,,& \qquad n = 2\,,\\[5pt]
			R = 0.4\,,& \qquad E_T^\mathrm{thres} = 4.8~\mathrm{GeV}\,,& \qquad \varepsilon = 0.0042\,,
			\end{array}
		\end{equation}
	where the fixed-cone parameters ($R,E_T^\mathrm{thres},\varepsilon$) correspond to the isolation set-up used by ATLAS.
	
	The theory prediction for $R_{13/8}^\gamma$ has not been performed as an independent calculation, but rather has been derived using the two calculations for $8~\mathrm{TeV}$ and $13~\mathrm{TeV}$. For both the theoretical uncertainty is estimated by means of a seven-point scale variation, $\mu_F=a\,p_T^\gamma$, $\mu_R=b\,p_T^\gamma$ with $a,b\in\{1/2,1,2\}$, where we exclude the configurations with $a/b\in\{1/4,4\}$.

	The uncertainty of $R_{13/8}^\gamma$ has now been estimated by forming the ratio for all possible combinations of the seven scale configurations for numerator and denominator, excluding again the combinations where the ratio of any two scales equals $1/4$ or $4$. This effectively corresponds to a generalisation of the seven-point scale variation for two scales to a 31-point variation for four scales.
	
	In figure~\ref{fig:Ratio} we show the result in the four rapidity bins mentioned above and compare to ATLAS data~\cite{ATLAS2019}. Except for the highest bins in $p_T^\gamma$ the description of the data is excellent. Like for the calculations~\cite{paper} for individual $\sqrt{s}$ we see a significant reduction in the uncertainty when going from NLO to NNLO: While at NLO the uncertainty lies between $(+10,-9)\%$ and $(+17,-14)\%$, only slightly growing with $p_T^\gamma$ and $|y^\gamma|$, at NNLO it lies between $(+3.4,-2.8)\%$ and $(+6.5,-4.0)\%$.


\section*{References}

\begin{thebibliography}{99}

\bibitem{ATLAS}The ATLAS Collaboration, \Journal{\JHEP}{08}{005}{2016}, \Journal{\PLB}{B770}{473-493}{2017}

\bibitem{ATLAS2}The ATLAS Collaboration, \Journal{\PLB}{B780}{578-602}{2018}

\bibitem{CMS}The CMS Collaboration, \Journal{\JHEP}{10}{128}{2015}, \Journal{\EPJ}{C79}{20}{2019}

\bibitem{MCFM}J.M. Campbell {\it et al}, \Journal{\PRL}{118}{222001}{2017}, \Journal{\PRD}{D96}{014037}{2017}

\bibitem{paper}X. Chen {\it et al}, arXiv:1904.01044 

\bibitem{Antenna} A. Gehrmann-De Ridder {\it et al}, \Journal{\JHEP}{09}{056}{2005}, A. Daleo {\it et al} \Journal{\JHEP}{04}{016}{2007}, J. Currie {\it et al} \Journal{\JHEP}{04}{066}{2013}

\bibitem{Frixione} S. Frixione, \Journal{\PLB}{B429}{369-374}{1998}

\bibitem{Hybrid} F. Siegert, \Journal{\em J. Phys.}{G44}{044007}{2017}

\bibitem{ATLAS2019}The ATLAS Collaboration, \Journal{\JHEP}{04}{093}{2019}

\bibitem{NNPDF}The NNPDF Collaboration, \Journal{\EPJ}{C77}{663}{2017}

\end{thebibliography}
\end{document}